\documentclass[preprint, double-spaced]{revtex4}
\usepackage[colorlinks=true,linktocpage=true,linkcolor=blue,citecolor=blue]{hyperref}
\usepackage{graphicx}
\usepackage{amsmath,bbm}
\usepackage{amssymb,bm}
\usepackage{array}

\begin{document}
\title{Multiplicity and Pseudorapidity distributions of charged particles in asymmetric and deformed nuclear collisions in a Wounded Quark Model}
\author{O. S. K. Chaturvedi$^1$}
\author{P. K. Srivastava$^{1,}$\footnote{prasu111@gmail.com}}
\author{Ashwini Kumar$^{2,}$\footnote{ashwini.physics@gmail.com}}
\author{B. K. Singh$^{1,}$\footnote{bksingh@bhu.ac.in}}
\affiliation{$^1$Department of Physics, Institute of Science, Banaras Hindu University, Varanasi-221005, INDIA}
%\affiliation{$^2$Department of Physics, Indian Institute of Technology Roorkee, Roorkee- 247667, INDIA}
\affiliation{$^2$School of Physical Sciences, National Institute of Science Education and Research, Bhubaneswar- 751005, INDIA}
\begin{abstract}
%% Text of abstract
The charged particle multiplicity ($n_{ch}$) and pseudorapidity density $(dn_{ch}/d\eta)$ are key observables to characterize the properties of matter created in heavy ion collisions. The dependence of these observables on collision energy and the collision geometry are a key tool to understand the underlying particle production mechanism. Recently a lot of focus has been made on asymmetric and deformed nuclei collisions since these collisions can provide a deeper understanding about the nature of quantum chromodynamics (QCD). On phenomenological perspective a unified model which describes the experimental data coming from various kind of collision experiments, is much needed to provide the physical insights about the production mechanism. In this paper, we have calculated the charged hadron multiplicities for nucleon-nucleus (such as proton-lead $(p-Pb)$ and asymmetric nuclei collisions like deutron-gold $(d-Au)$, and copper-gold $(Cu-Au)$ within a new version of wounded quark model (WQM) and shown their variation with respect to centrality. Further we have used a suitable density function within our WQM to calculate pseudorapidity density of charged hadrons at mid-rapidity in the collisions of deformed uranium nuclei. We found that our model with suitable density functions describes the experimental data for symmetric, asymmetric and deformed nuclei collisions simultaneously over a wide range of collision energy.
\end{abstract}

\maketitle 
\section{Introduction}
\noindent
The multiparticle production in ultra-relativistic heavy ion collisions is an important tool to study the perturbative as well as non-perturbative nature of Quantum Chromo dynamics (QCD)~\cite{1,2,3}. Enormous data has been collected by Relativistic Heavy Ion Collider (RHIC) and Large Hadron Collider (LHC) from various types of collision e.g. gold-gold $(Au-Au)$, lead-lead $(Pb-Pb)$, deutron-gold $(d-Au)$, copper-gold $(Cu-Au)$, uranium-uranium $(U-U)$, proton-lead $(p-Pb)$ etc. over a wide range of center-of-mass energy~\cite{4,5,6,7,8,9,10,11,12,13,14,15,16,17,18,19}. The charged particle multiplicity ($n_{ch}$) and their pseudorapidity density ($dn_{ch}/d\eta$) at central rapidity are fundamental quantities measurable in high energy experiment which serve as an important tools to analyze these data and characterize the global properties of the systems created in the heavy ion collision. Due to variation in shape, size and orientation of the colliding nuclei, the collisions can have different kind of initial geometries. The dependence of the $n_{ch}$ and $(dn_{ch}/d\eta)_{\eta=0}$ on the collision geometry is sensitive to the underlying particle production mechanism. These distributions can be used to quantify the contribution of soft and hard processes in particle generation, stopping and penetrating powers of the colliding nuclei, contribution ratios of leading nucleons, square of speed of sound etc.~\cite{23}. 
 
 The particle production in heavy ion collisions is quite involved process since the colliding nuclei in these collisions do not behave as a mere incoherent superposition of their constituent nucleons. Rather, coherence effects become important and they modify not only the partonic flux into the collision, but also the underlying dynamics of particle production in the scattering processes. Therefore we need some baseline nucleon-nucleon collision experiments e.g. proton-proton $(p-p)$, proton-antiproton ($p-{\bar p}$) etc. and also the nucleon-nucleus (or some very small nucleus colliding with a large nucleus) collision experiments. Data from baseline experiments are useful to understand the particle production mechanism in vacuum and nucleon-nucleus $(n-A)$ collisions to study the pure nuclear effects (such as shadowing, anti-shadowing, absorption, saturation etc.) on particle production mechanism. The data from these collisions also helps to decipher the initial and final state effects for a proper characterization, as these effects may lead to qualitatively similar phenomena in observables. 

On the other side asymmetric and deformed nuclei collisions e.g., $Cu-Au$ collisions, $U-U$ collisions etc. are interesting to study due to the different initial geometrical configurations of the colliding systems~\cite{17}. For example a substantial initial geometrical asymmetry which existed in $Cu-Au$ collisions could lead to naturally arising odd harmonics, from both the core and/or the corona~\cite{19}. In peripheral $Cu-Au$ collisions a sizable directed flow from $Au$ to $Cu$ nucleus is generated due to initial asymmetry of electric charges on two nuclei and thus can be useful to study the electromagnetic and chiral magnetic properties of quark gluon plasma~\cite{25}. Since $Cu$ nucleus is completely swallowed by the $Au$ in central collisions, it could also provide an opportunity to measure an extensive range of initial energy densities for this system. Recently an analysis made by PHENIX collaboration shows that the hadron multiplicities in $Cu-Au$ collisions show the same centrality trend as observed in $Au-Au$ collisions at $200$ GeV center-of-mass energy. However the transverse energy calculated for $Cu-Au$ is higher than the $Au-Au$ system which is contrary to earlier expectations~\cite{26}. Similarly the study of deformed uranium-uranium $(U-U)$ collisions at RHIC has gained a lot of interest in past few years. The unique geometry and shape of uranium nucleus provide opportunities to understand the particle production mechanism and elliptic flow~\cite{27,28}. Uranium nucleus is not spherical and has a prolate shape, which leads to different collision geometry shapes from body on body to tip on tip configurations~\cite{29,30}, even in fully overlap collisions. In the case of tip on tip collisions, the general expectation was that the produced particle multiplicity should be higher due to larger number of binary collisions, while the elliptic flow are smaller since the overlap region is symmetric, while in case of body on body collisions smaller multiplicity associated with larger elliptic flow. However recent data of $U-U$ collisions for most central $1$$\%$ event from PHENIX experiment show almost no enhancement of multiplicity in comparison to symmetric $Au-Au$ collisions~\cite{17}. These results can only be understood with the help of theoretical models. These models can only provide the physical insights to the underlying particle production mechanism. A lot of effort has been put forward in this direction~\cite{28,29,30,31,32,33,34,35,36,37,38,39,40,41,42,43,44,45,46,47,48,49,50,52,53}. However, a phenomenological model which describe these multiplicity distribution data simultaneously for various types of collision is still needed. Recently we have proposed a parametrization which is based on a phenomenological model involving wounded quarks interactions~\cite{70,71,72}. Our model with minimal number of parameters successfully explains the charged hadron multiplicity distributions for symmetric nucleus-nucleus collisions such as $Cu-Cu$, $Au-Au$ and $Pb-Pb$ etc~\cite{73,74}. In this paper our main aim is to extend our model suitably so that we can accommodate asymmetric and deformed nuclei collisions along with symmetric collisions. Here we first improve our parametrization for nucleon-nucleon collisions by including recent experimental data and then we extend our WQM by changing its density profile function to describe the various type of collisions having different initial geometrical configurations. 

Rest of the paper is organised as follows: In Section II, firstly we provide a parametrization for the total multiplicity and pseudorapidity density at mid-rapidity of charged hadrons which produce in $p-p$ collisions. Section III will provide a brief description of WQM. Here we also present the different nuclear density functions for  symmetric, asymmetric and deformed nuclei. In section IV we show the results obtained in our model and their comparison with the corresponding experimental data wherever available. Finally we summarize our present study.
\section{Parametrization for p-p Collisions}
\begin{figure}
\includegraphics[scale=0.55]{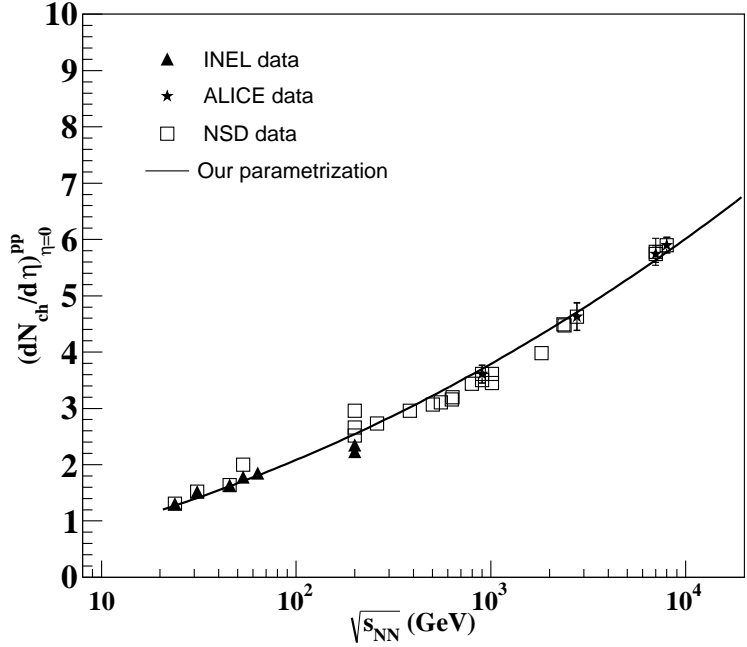}
\caption{Variation of $(dn_{ch}/d\eta)_{\eta=0}$ produced in $p-p$ interaction with respect to center of mass energy ($\sqrt{s_{NN}}$). Filled symbols are data from inelastic $p-p$ events and open symbols are experimental data for NSD events~\cite{54,55,56,57,58,59,60,61,62,63,64,65,66,67}. A fitting using the parametrization as written in Eq.(2) is shown. ALICE data is taken from reference~\cite{4}.}
\end{figure}
The production of charged hadrons in nucleus-nucleus collisions are deeply linked with $p-p$ collisions at various energies. Initially, Feynman pointed out that the charged hadron multiplicity distribution in $p-p$ collisions have no dependence on the available energy as centre-of-mass energy ($\sqrt{s}) \to \infty$. This implies that total charged hadron multiplicity after summing over rapidity have $ln{\sqrt{s}}$ dependence since $y_{max}=ln{\frac{\sqrt{s}}{m_{N}}}$, where  $m_N$ is the nucleon mass. Further additional gluons arising from gluon-bremsstrahlung processes gives QCD radiative corrections to the total charged hadron production, hence total charged hadron multiplicity have $ln^{2}\sqrt{s}$ dependence. Recent data regarding pseudorapidity distributions of charged hadrons coming from PHOBOS experiment for $p-p$ and $p-\bar{p}$ in central plateau region i.e  $(\frac{dn}{d\eta})_{\eta=0}$ shows $ln^{2}\sqrt{s}$ dependence which will finally give $ln^{3}\sqrt{s}$ type dependence to the total charged hadron multiplicity distribution~\cite{20,21,22}..
Based on these experimental findings and exploiting the QCD hypothesis of universal particle production in various hadronic collisions, we have proposed a parametrization involving a cubic logarithmic term to calculate the charged hadron mean multiplicity in $p-p$ collisions at any collision energy as follows:
%We have used the parametrization with same parameters value as used in our earlier papers~\cite{45,46} for total mean charged hadron multiplicity in p-p collisions : 
\begin{equation}
 <n_{ch}> _{pp}=(a'+b' ln \sqrt{s_{a}}+c'ln^{2} \sqrt{s_{a}}+d'ln^{3} \sqrt{s_{a}})-\alpha.
\end{equation}
In Eq. (1), $\alpha$ is the leading particle effect which arises due to the energy carried away by the spectator quarks and its value is experimentally determined as 0.85. $\sqrt{s_{a}}$ is the available center-of-mass energy i.e., $\sqrt{s_{a}}=\sqrt{s}-m_{B}-m_{T}$, where $m_{B}$ is the mass of projectile and $m_{T}$ the mass of the target nucleon, respectively; $a'$, $b'$, $c'$ and $d'$ are constants having the  values $a'=1.8$, $b'=0.37$, $c'=0.43$ and $d'=0.04$~\cite{45,46}.
To understand particle production mechanism from high energy $h-h$ to $A-B$ collisions, pseudorapidity distribution is very useful quantity. It was pointed out that $\frac{dn_{ch}}{d\eta}$ is suitably used to find the information about the temperature and energy density of QGP. For pseudorapidity density at mid-rapidity, we have proposed a parameterization in our earlier publications \cite{71,72}. Here we use the same parametrization but do a reasonable fit again to accommodate the new data come from $p-p$ collision data at $0.9$, $1.8$, $7$ TeV energies~\cite{4}:
\begin{equation}
 <(dn_{ch}/d\eta)^{pp}_{\eta=0}>=(a_{1}^{'}+b_{1}^{'} ln \sqrt{s_{a}}+c_{1}^{'}ln^{2} \sqrt{s_{a}})-{\alpha_1}'.
\end{equation}
We have obtained slightly changed values of the parameters as compared to the values obtained in our earlier publications \cite{73,74}. The values are $a_{1}' = 1.15$, $b_{1}' = 0.16$, and $c_{1}' = 0.05$.
\section{Model Formalism}
Extrapolating our parametrization from $p-p$ collisions to $h-A$ collisions we have assumed that the quark-quark picture is more suitable than nucleon-nucleon picture for determining charged hadron multiplicity distribution. We consider here that a wounded quark which participate in the reaction suffered multiple collisions before it hadronizes. Based on this, we have used the expression for average charged hadron multiplicity in $h-A$ collisions as follows ~\cite{73,74}:
\begin{equation}
 <n_{ch}> _{hA}=N_{q}\left[a'+b' ln \left(\frac{\sqrt{s_{A}}}{N_{q}}\right)+c'ln^{2}\left(\frac{\sqrt{s_{A}}}{N_{q}}\right) +d'ln^{3}\left(\frac{\sqrt{s_{A}}}{N_{q}}\right) \right]-\alpha.
\end{equation}
In Eq. (3), $\sqrt{s_{A}}$  is related to  ${s_{A}}$ as  $\sqrt{s_{A}}$ = $\nu_{qA}$ ${s_{A}}$ where $\nu_{qA}$ represents the mean number of collisions of the wounded quark inside the nucleus $A$ and is defined by,
\begin{equation}
\nu_{qA} = \frac{A\sigma_{qN}^{in}}{\sigma_{qA}^{in}}.
\end{equation}
Here $\sigma_{qN}^{in}$ and $\sigma_{qA}^{in}$ are the inelastic cross-sections for quark-nucleon $(q-N)$ and quark-nucleus $(q-A)$ interactions, respectively and $A$ is the atomic mass of the target nucleus. 
 In the above equation the mean number of constituent quarks which becomes wounded in $h-A$ collisions shares the total available centre-of-mass energy $\sqrt{s_{A}}$ according to law of equipartition of energy. Due to this the energy available to each interacting quarks becomes $\sqrt{s_{A}}/N_q$. The parameters $a$, $b$, $c$ and $d$ do not change its value in $h-A$ collisions as compared to $p-p$ collisions and thus gives us hint that they are directly related to quark-quark and quark-gluon interaction processes in QCD. In earlier publications by Singh et al. \cite{70,71,72}, the importance of the parameter $c$ is shown where it comes automatically from gluon-bremsstrahlung process in QCD. Further in Eq. (3) $N_{q}$ is the mean number of inelastically interacting quarks with the nuclear targets and defined as,
\begin{equation}
N_{q}^{hA} = \frac{N_{c}\sigma_{qA}^{in}}{\sigma_{hA}^{in}}.
\end{equation}
Here, $\sigma_{qA}^{in}$ and $\sigma^{in}_{hA}$ are the scattering cross sections for quark-nucleus and hadron-nucleus interactions, respectively obtained from Glauber's theory.
The quark-nucleus inelastic interaction cross-section $\sigma_{qA}^{in}$ is determined from $\sigma_{qN}^{in}(= 1/3\sigma_{NN}^{in})$ by using Glauber's approximation as follows :
\begin{equation}
\sigma_{qA}^{in}=\int d^{2}b\left[1-\left(1-\sigma_{qN}^{in}D_{A}(b)\right)^{A}\right],
\end{equation}
where profile function $D_{A}(b)$ is related to nuclear density, $\rho(b,z)$ by the relation :
\begin{equation}
D_{A}(b)=\int_{-\infty}^{\infty}\rho(b,z)dz.
\end{equation}
The fireball size formed in the hadronic and nuclear collisions depends on the overlap cross-section of colliding systems; which indirectly results in the change in the mean number of participating quarks as we move towards peripheral collisions. $N_{q}$ becomes maximum in central collisions and It decreases when we move to peripheral collisions. This fact shows $N_{q}$ has a centrality as well as colliding systems dependence. \\   
Since we are discussing the collisions happen among small-large nuclei, deformed-deformed nuclei along with large-large nuclei so we have to choose an appropriate density functions to properly describe the charged density of different nuclei. We have used the following functions :
\subsection{For large nuclei}
We have used the Woods-Saxon charge distribution for large nuclei which can be expressed as follows:
\begin{equation}
\rho(b,z)=\frac{\rho_{0}}{1-exp(\frac{\sqrt{b^{2}+z^{2}}-R}{a})},
\end{equation}
where all the notations have their usual meaning~\cite{73,74} .
\subsection{For small deuteron nuclei}
We have used following Hulth\'en function for expressing charge density of deuterium nuclei~\cite{68,69} 
\begin{equation}
  \rho(r)=\rho_0 \left(\frac{e^{-ar}+e^{-br}}{r}\right)^2,
\end{equation}
 where $a=0.457$~fm$^{-1}$ and $b=2.35$~fm$^{-1}$.

\subsection{For deformed nuclei}
For deformed nuclei we have used the following modified form of Woods-Saxon charge distribution~\cite{68,69}: 
\begin{equation}\label{eq:deformed}
  \rho(x,y,z)=\rho_0 \frac{1}{1+\exp\frac{\left(r-R(1+\beta_2 Y_{20} +\beta_4 Y_{40})\right)}{a}}\,,
\end{equation} 
where $Y_{20}=\sqrt{\frac{5}{16\pi}}({\rm cos}^2(\theta)-1)$, 
$Y_{40}=\frac{3}{16\sqrt{\pi}}(35{\rm cos}^4(\theta)-30 {\rm cos}^2(\theta)+3)$ are the spherical harmonics with the
deformation parameters $\beta_2$ and $\beta_4$. 

The different parameters value for uranium nuclei is taken from Refs. ~\cite{68,69}. 
\\
The generalization of the the $h-A$ to nucleus-nucleus collisions goes along the same line in WQM and can be given as follows~\cite{73,74} :
\begin{equation}
<n_{ch}> _{AB}=N_{q}^{AB}\left[a'+b' ln \left(\frac{\sqrt{s_{AB}}}{N_{q}^{AB}}\right)+c'ln^{2}\left(\frac{\sqrt{s_{AB}}}{N_{q}^{AB}}\right) +d'ln^{3}\left(\frac{\sqrt{s_{AB}}}{N_{q}^{AB}}\right) -\alpha\right],
\end{equation}
where $\sqrt{s_{AB}}=A(\nu_{q}^{AB}s_{a})^{1/2}$ where $A$ is the mass number of colliding nucleus and the mean number of inelastic quark collision $\nu_{q}^{AB}$ can be given as follows :
\begin{equation}
\nu_{q}^{AB}=\nu_{qA}\nu_{qB}=\frac{A\sigma_{qN}^{in}}{{\sigma_{qA}^{in}}}.\frac{B\sigma_{qN}^{in}}{{\sigma_{qB}^{in}}}.
\end{equation}
Furthermore, mean number of participating quarks $N^{AB}_{q}$ can be calculated by generalizing Eq. (9) in the following manner:
\begin{equation}
N^{AB}_{q}=\frac{1}{2}\left[\frac{N_{B}\sigma_{qA}^{in}}{{\sigma_{AB}^{in}}}+\frac{N_{A}\sigma_{qB}^{in}}{{\sigma_{AB}^{in}}}\right],
\end{equation}
where $\sigma_{AB}^{in}$ is the inelastic cross-section for $A-B$ collision and can be expressed in the following manner:
\begin{equation}
\sigma_{AB}^{in}= \pi r^{2}\left[A^{1/3}+B^{1/3}-\frac{c}{A^{1/3}+B^{1/3}}\right]^2,
\end{equation}
where $c$ is a constant and has a value $4.45$ for nucleus-nucleus collisions. 

Now, for central average charge hadron multiplicity can be found by following way:
\begin{equation}
<n_{ch}>^{central}_{AB}=A\left[a'+b'ln(\nu_{q}^{AB}s_{a})^{1/2}+c'ln^{2}(\nu_{q}^{AB}s_{a})^{1/2}+d'ln^{3}(\nu_{q}^{AB}s_{a})^{1/2}-\alpha\right].
\end{equation}
Now, we want to provide an expression to calculate pseudorapidity distribution of charged hadrons with respect to pseudorapidity. For this we take the help of wounded nucleon two component model.
The physical interpretation of usual two-component model based on hadron production by longitudinal projectile nucleon dissociation (soft component) and transverse large-angle scattered parton fragmentation (hard component). However, here in our wounded quark picture the hard component scales with the number of quark-quark collisions and the soft component scales with the participant quarks number. Thus in $A-B$ collisions central mean pseudorapidity density can be parametrized in terms of $p-p$ rapidity density as follows \cite{73,74}: 
\begin{equation}
\left(\frac{dn_{ch}}{d\eta}\right)^{AA}_{\eta=0}=\left(\frac{dn_{ch}}{d\eta}\right)^{pp}_{\eta=0}\left[\left(1-x\right)N_{q}^{AB}+ x N_{q}^{AB}\nu_{q}^{AB}\right],
\end{equation}
where $(\frac{dn_{ch}}{d\eta})^{pp}_{\eta=0}$ is calculated from Eq. (2) using the new parameter values.

where $x$ signifies the relative contributions of hard and soft processes in two component model. The value of $x$ varies from $0.1$ to $0.125$ with centre-of-mass energy. 
\section{Results and Discussions}
\subsection{Total Multiplicity and Pseudorapidity density}
In earlier publication we have shown that WQM provides a suitable description to the various features of charged hadron production in high energy collisions such as multiplicity distributions with respect to pseudorpidity and collision energy for symmetric collisions of nuclei with varying sizes~\cite{73}. Recently the multiplicity data for symmetric $Pb-Pb$ collisions at 2.76 TeV and 5.02 TeV has been presented by LHC experiment. Thus we first want to show our result corresponding to $n_{ch}$ and $dn_{ch}/d\eta$ for this symmetric collision and then move towards asymmetric and deformed nuclei collisions. In Fig. 2 we have shown the variation of $\sigma_{qPb}^{in}$ with respect to transverse size of the collision zone for $Pb-Pb$ collision at $2.76$ TeV. Here we have shown the selection criteria in present model for different centrality bin using a cut on $\sigma_{qPb}^{in}$. We have used this centrality criteria to calculate the charged hadron multiplicity with respect to centrality for $Pb-Pb$ collisions at $2.76$ TeV in Fig. 3. Further we have compared our results with the corresponding experimental data. We find that the model suitably describes the data. We have also shown WQM results and its comparison with experimental data in tabular form in Table I for four centrality classes.  
\begin{figure}
\includegraphics[scale=0.55]{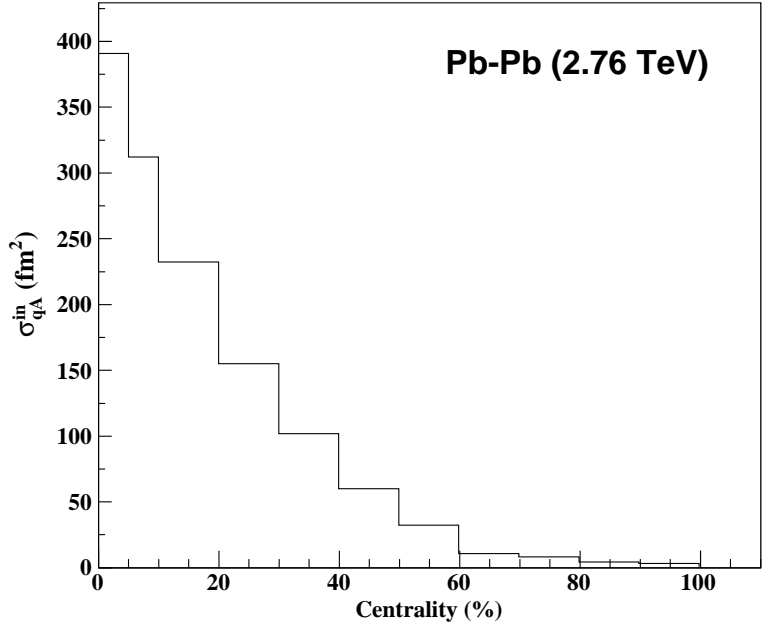}
\caption{Variation of the quark-nucleus inelastic cross-section ($\sigma_{qA}^{in}$) in our model as a function of centrality for $Pb-Pb$ collisions $\sqrt{s_{NN}}$ = $2.76$ TeV at LHC energy.}
\end{figure}

\begin{figure}
\includegraphics[scale=0.55]{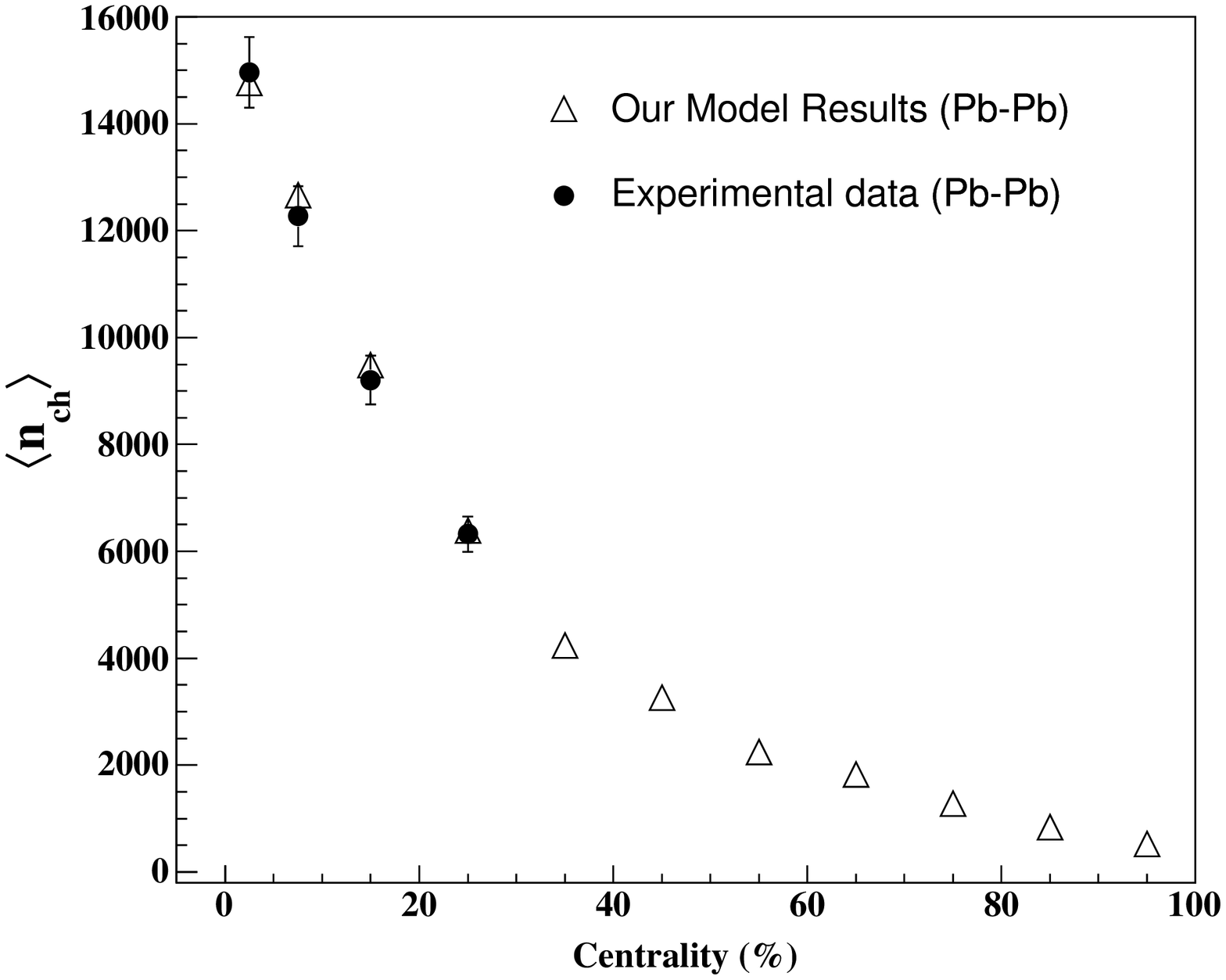}
\caption{Variation of mean charged hadron multiplicity with respect to centrality for $Pb-Pb$ collisions at $2.76$ TeV. Open triangles are the WQM results and solid circles are the experimental data taken form Ref.~\cite{5}.}
\end{figure}

Rapidity density of charged particles is quite useful in providing a qualitative measure of the final entropy per unit rapidity produced in a collision. It depends essentially on system size, centrality of the collision event and collision energy. Additionally, It also gives an indication towards the viscous effect during expansion of hot and dense QCD matter created in a collision as viscous effects contributes to the entropy generation and in turn, enhance the final charged particle multiplicity. In Fig. 4 we have presented the variation of charged hadron pseudorapidity density at midrapidity with respect to centrality for $Pb-Pb$ collisions at 2.76 TeV. Further, we have shown our model results with the experimental data along with other model results for a comparative study. We have shown here the results obtained from HIJING~\cite{24}, AMPT~\cite{24} with different sets of parameter along with the the results obtained from recent version of DPMJET~\cite{39}. Here we find that most of these models are able to satisfy the data obtained in peripheral collisions. However, for semi-peripheral and central collisions, a varying level of agreement is observed as some of these models overpredict and some of these underpredict the data. Specially the difference between result obtained from HIJING with jet quenching on and the experimental data is quite large in central collisions. On the other hand, WQM results satisfy the experimental data quite well in consistent manner from central to peripheral collisions.  

In Fig. 5, we have plotted the variation of pseudorapidity density of charged hadrons at midrapidity divided by number of participating quarks with respect to centrality for $Pb$-$Pb$ and $Au-Au$ collisions at $\sqrt{s_{NN}} = 2.76$ TeV and $200$ GeV, respectively. We have shown the corresponding $p-p$ data in the most peripheral centrality class of $Pb-Pb$ and $Au-Au$ collisions since the mean  number of participating quarks for $p-p$ collision is 2 which is almost equal to the number of participating quarks in $Pb-Pb$ and $Au-Au$ collision in their most peripheral collisions. One can observe from the graph that $(dn_{ch}/d\eta)_{\eta=0}$ divided by number of participating quarks is almost independent of centrality which is actually the basic theme of wounded quark picture. Some deviation is due to different transverse momentum or different entropy production at different centralities during the evolution of the system~\cite{41,42}. Further, the mean number of charged hadrons per unit pseudo-rapidity coming from a participating wounded quark pair is same within error bars for $Pb$-$Pb$ and $p$-$p$ collisions both. We have also found this scaling feature in $Au-Au$  and $p-p$ collisions at RHIC energy. These observations support the idea of independent particle production from a wounded quark source.
  
\begin{figure}
\includegraphics[scale=0.55]{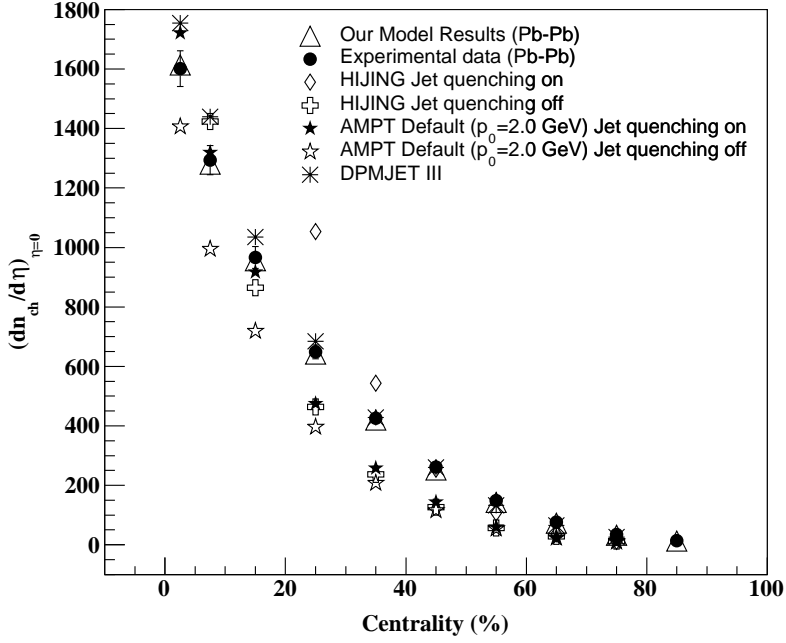}
\caption{Variation of pseudorapidity density with respect to centrality for $Pb-Pb$ collisions at $2.76$ TeV~\cite{6}.}
\end{figure}

\begin{figure}
\includegraphics[scale=0.55]{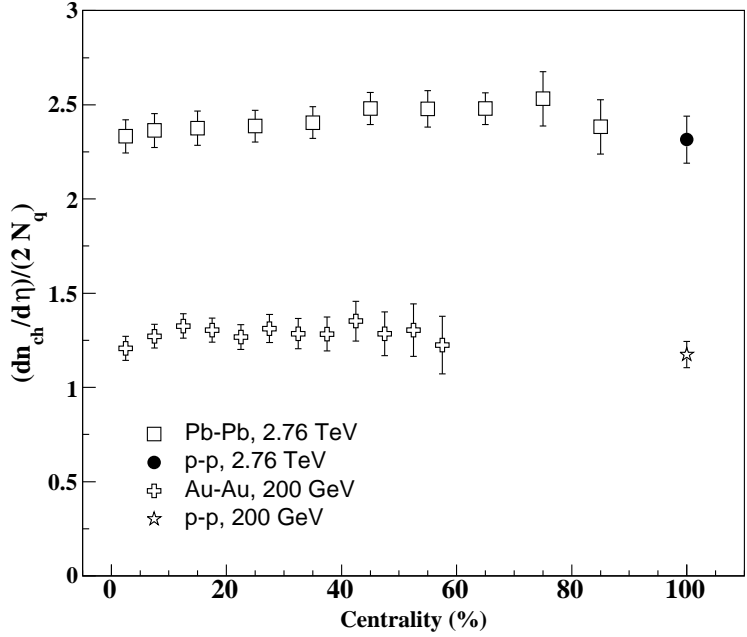}
\caption{Variation of $(dn_{ch}/d\eta)/(2 N_{q})$ as a function of centrality for $Pb-Pb$ and $Au-Au$ collisions at $\sqrt{s_{NN}}$ = $2.76$ TeV and $200$ GeV, respectively. We have also shown $(dn_{ch}/d\eta)/(2 N_{q})$ for $p-p$ collisions at corresponding energies. Here $N_{q}$ is calculated using WQM and experimental data of $(dn_{ch}/d\eta)$ is taken from Refs. ~\cite{4,6,12,62}.}
\end{figure}

\begin{table*}
\begin{center}
\caption{The total charge hadron multiplicities as a function of centrality in $Pb-Pb$ Collisions LHC energy. The data is shown here is taken from ALICE experiment~\cite{5}.}
\begin{tabular}{|l|l|l|}\hline
{Centrality Bin} & \multicolumn{2}{l|}{{$\sqrt{s_{NN}}=2.76$ TeV}}  \\
\cline{2-3}

 &{Model}&{Experimental }\\
\hline\hline

 ~$0-5 \%$  & 14764       &14963$\pm$666     \\
 ~$5-10 \%$  & 12667       &12272$\pm$561                \\
$10-20 \%$   & 9484      &9205$\pm$457           \\
$20-30 \%$   & 6384        &6324$\pm$330                     \\ \hline
\end{tabular}
\end{center}
\end{table*}

\begin{figure}
\includegraphics[scale=0.55]{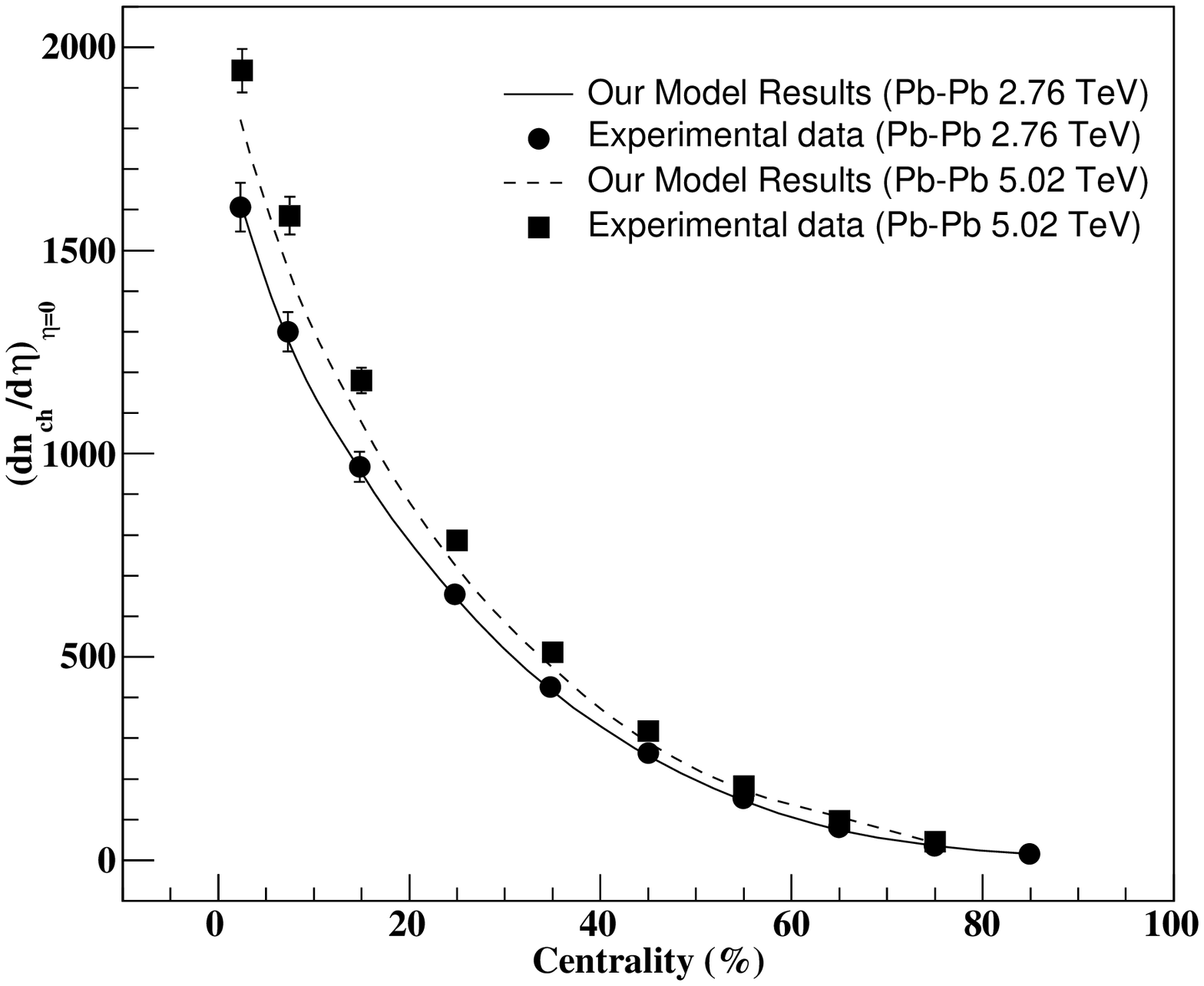}
\caption{Variation of pseudorapidity density at mid-rapidity with respect to centrality for $Pb-Pb$ collisions at $2.76$ TeV~\cite{6} and $5.02$ TeV~\cite{14}.}
\end{figure}

In Fig. 6, we have shown the model results for charged hadron pseudo-rapidity density at mid-rapidity at $2.76$ TeV and $5.02$ TeV and compare them  with the experimental data at the two LHC energies. Model and experimental results at the two LHC energies are show a good level of agreement. Further, the observed increase in the multiplicity at $5.02$ TeV with respect to the multiplicity at $2.76$ TeV is seen which is quite obvious due to the increase in the collision energy, resulting into the higher production at $5.02$ TeV.

In Fig. 7, we have presented the variation of  $\sigma_{qd}^{in}$.$\sigma_{qAu}^{in}$ with respect to transverse coordinate of the fireball formed in d-Au collision to determine the different centrality class. Fig. 8 presents the variation of charged hadron multiplicity with respect to centrality for $d-Au$ collisions. We have used Hulthen density function to describe the charged distribution in deuteron nucleus and Woods-Saxon density function for gold nucleus. We compare our model result with the corresponding experimental data from PHENIX experiment and found a good agreement between them.  
\begin{figure}
\includegraphics[scale=0.55]{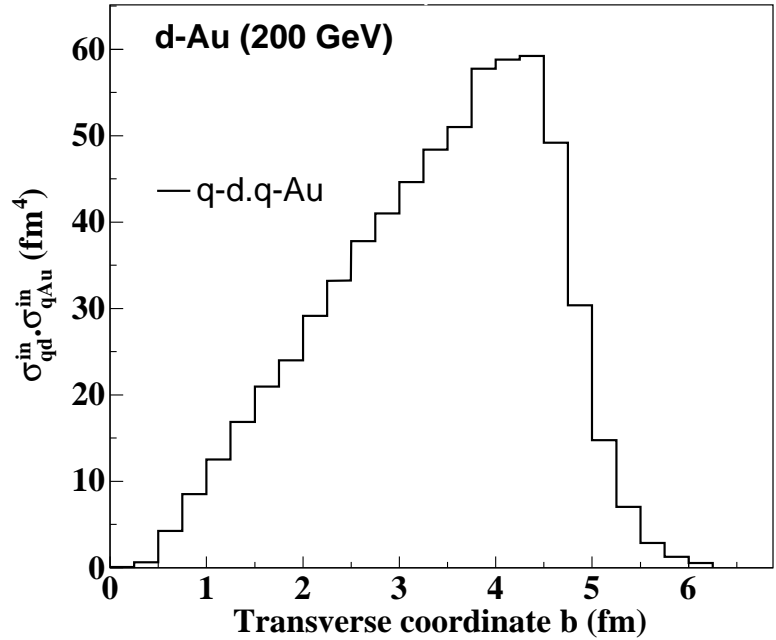}
\caption{Variation of $\sigma_{qd}^{in}$.$\sigma_{qAu}^{in}$ in our model as a function of the transverse coordinate $b$ (fm) of collision zone formed in $d-Au$ collisions at $200$ GeV.}
\end{figure}

\begin{figure}
\includegraphics[scale=0.55]{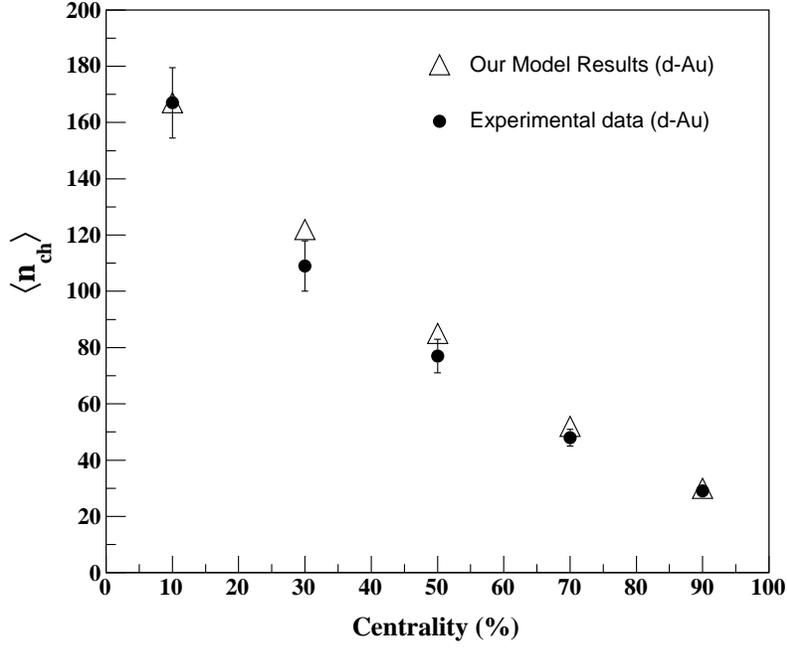}
\caption{Variation of total charge hadron multiplicity with respect to centrality for $d-Au$ collisions at $200$ GeV~\cite{9}.}
\end{figure}
\begin{figure}
\includegraphics[scale=0.55]{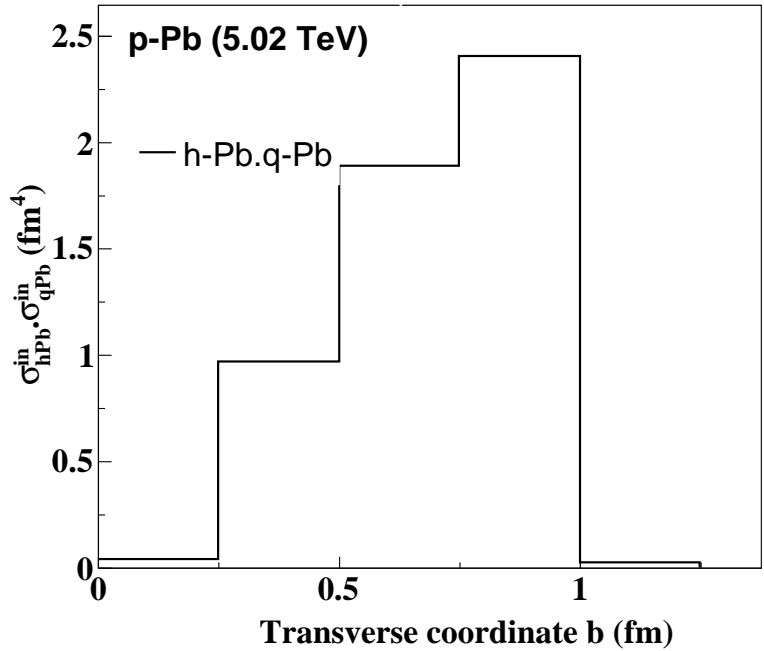}
\caption{Variation of $\sigma_{hPb}^{in}$.$\sigma_{qPb}^{in}$ in our model as a function of the transverse coordinate $b$ (fm) of collision zone formed in $p-Pb$ collisions at $5.02$ TeV.}
\end{figure}
\begin{figure}
\includegraphics[scale=0.55]{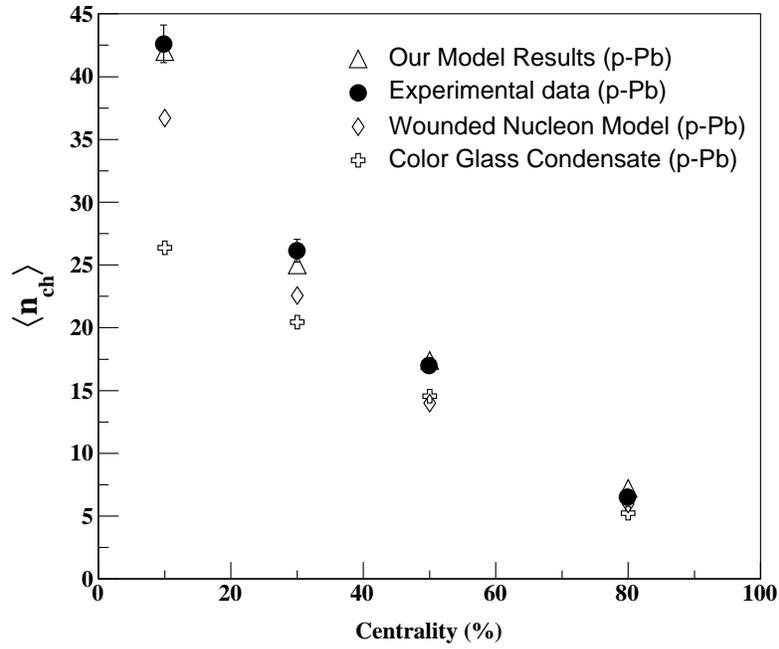}
\caption{Variation of total mean multiplicity with respect to centrality for $p-Pb$ collision at $5.02$ TeV~\cite{10}}
\end{figure}

\begin{figure}
\includegraphics[scale=0.55]{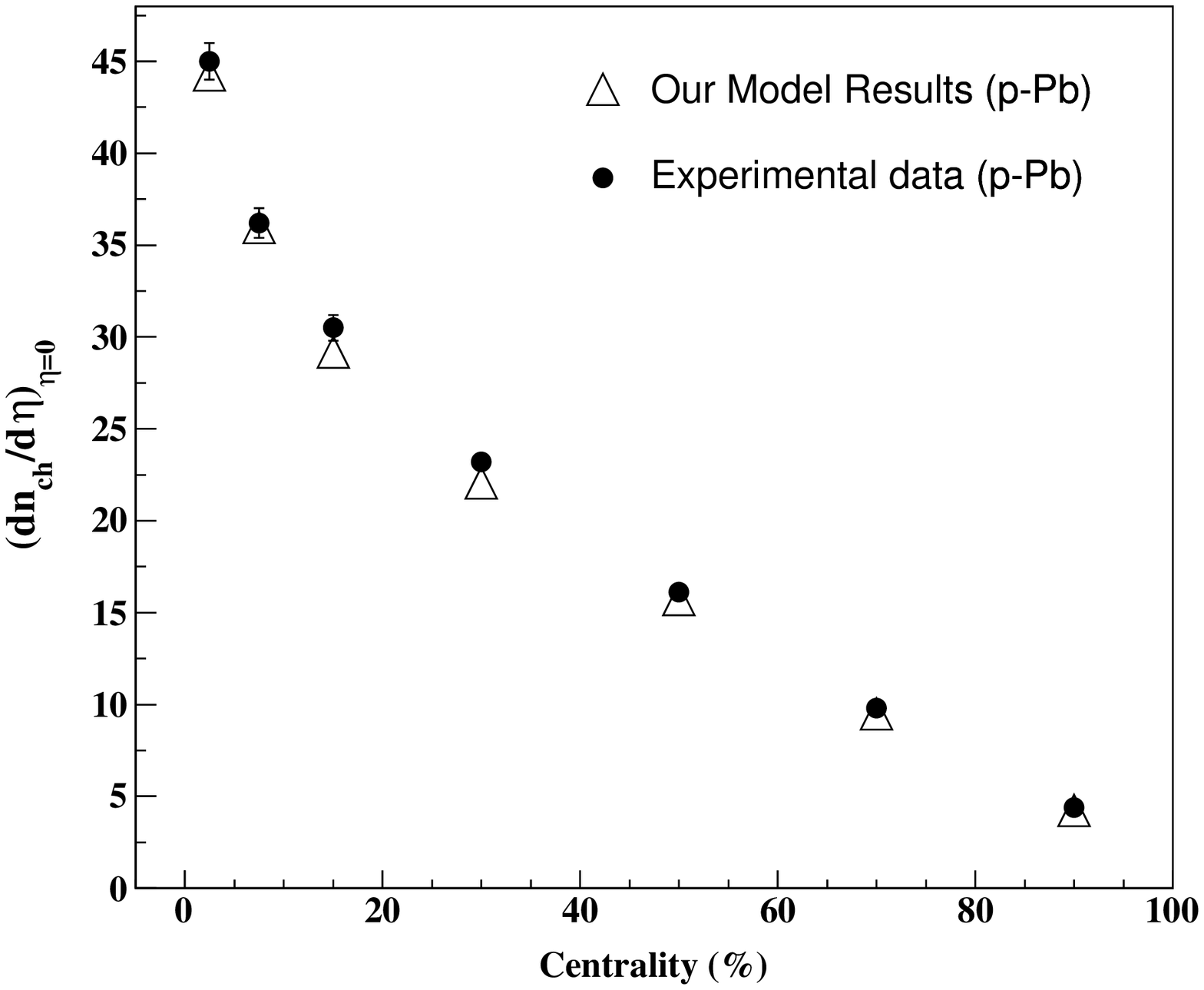}
\caption{Variation of pseudorapidity density at midrapidity with respect to centrality for $p-Pb$ collision at $5.02$ TeV~\cite{15}}
\end{figure}

 Fig. 9 represents the product of cross-sections of proton-lead collision and quark-lead collision with respect to transverse coordinate $b$ to define the various centrality class in $p-Pb$ collisions. In Fig. 10, the model results for the variation of mean multiplicity with respect to centrality is shown for $p-Pb$ collisions at $5.02$ TeV along with its comparison to experimental data at LHC energy. Further, we have shown the results from Wounded Nucleon Model and Color Glass Condensate for the sake of comparison. Our model results agree with the experimental data quite well whereas other model results show a varying level of agreement. The charged hadron multiplicity is $\approx$ $43$ for $p-Pb$ collisions in most central bin which is almost one-third of the multiplicity in most central bin of $d-Au$ collisions as tabulated in Table II, where charged hadron multiplicity is $167$. In spite of the big size of $Pb$-nucleus in comparison to $Au$, the multiplicity is low for $p-Pb$ case. Thus it suggests that a relatively larger fireball is expected to be formed in $d-A$ collision due to large overlap area than $p-Pb$ collision at higher energy $5.02$ TeV. Also, the number of collisions suffered by each wounded quark traveled through the medium is less in $p-Pb$ collisions as compared to the $d-Au$ collisions even at the cost of increased collisional energy. It clearly shows the significance of role played by collision geometry as smaller collision systems (such as $p-A$) will create less matter as compared to $d-A$ collision and will exist for a shorter span of time for particle production in final state. Similarly, in Fig. 11 we have shown the model results for pseudorapidity density at midrapidty with the centrality for $p-Pb$ collisions at $5.02$ TeV and a comparison is made with the available experimental data for the same energy. Again, It is noticeable that model results explains the experimental data quite well for all centrality bins. 

To define the centrality bin in asymmetric collisions like $Cu-Au$ is a bit complex process. In our model we have first plotted the $\sigma_{qCu}^{in}$ and $\sigma_{qAu}^{in}$ with respect to transverse size of the collision zone. After that we have plotted the product of both with respect to transverse size (as shown in Fig. 12) and then we divide the whole transverse fireball area into different centrality regions. Based on this analysis, we have calculated the model results for the variation of $dn_{ch}/d\eta$ with respect to centrality for asymmetric $Cu-Au$ collisions along with the experimental data for the comparative study (as shown in Fig. 13). Furthermore, we have shown comparison of WQM results with the corresponding result obtained in IP-Glasma model. We found that both the model i.e., WQM and IP-Glasma suitably describe the experimental data. 
\begin{figure}
\includegraphics[scale=0.55]{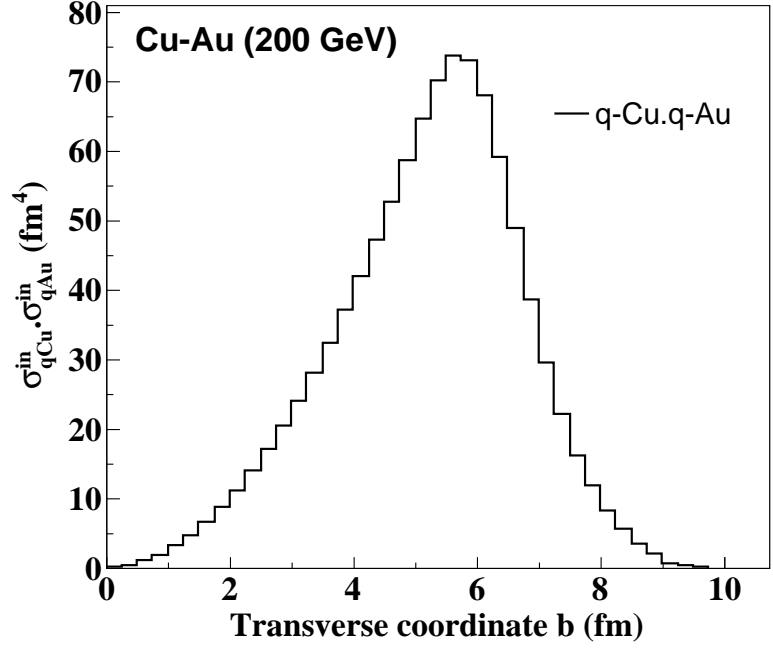}
\caption{Variation of $\sigma_{qCu}^{in}$.$\sigma_{qAu}^{in}$ in our model as a function of the transverse coordinate $b$ (fm) of collision zone formed in $Cu-Au$ collisions at $200$ GeV.}
\end{figure}

\begin{figure}
\includegraphics[scale=0.55]{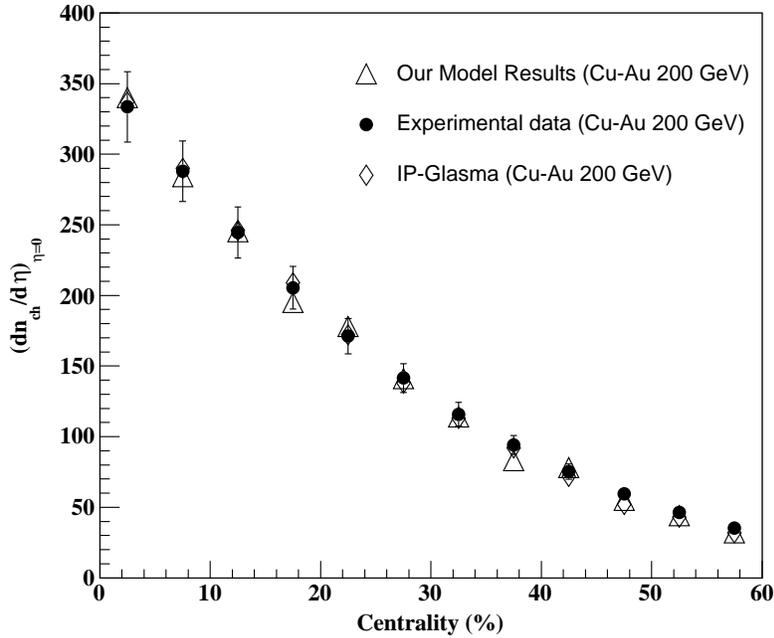}
\caption{Variation of pseudorapidity density at midrapidity with respect to centrality for $Cu-Au$ collision at $200$ GeV~\cite{12}.}
\end{figure}

\newpage

\begin{table*}
\begin{center}
\caption{The total charged hadron multiplicities as a function of centrality in $d-Au$ Collisions at RHIC energy. The experimental data has been taken from PHOBOS experiment~\cite{9}.}
\begin{tabular}{|l|l|l|}
\hline
{Centrality Bin} & \multicolumn{2}{l|}{{$\sqrt{s_{NN}}=200$ GeV}}  \\
\cline{2-3}
 &{Model}&{Experimental }\\
\hline\hline

  ~$0-20 \%$  & 167    & $167_{-11}^{+14}$             \\
 ~$20-40 \%$  & 122      & $109_{-8}^{+10}$                \\
$40-60 \%$   &  85      & $77_{-5}^{+7}$            \\
$60-80 \%$   &  52    & $48_{-3}^{+3}$               \\
$80-100 \%$   & 30    & $29_{-3}^{+3}$       \\ \hline
\end{tabular}
\end{center}
\end{table*}
\noindent

Now we move towards the deformed $U-U$ nuclei collisions. To describe the charge density function in $U$ nuclei, we have taken the deformed Woods-Saxon density function. For a particular centrality class there are different type of orientation configuration is possible for $U-U$ collisions. Measurement of charged particle multiplicity density in deformed nuclei is very sensitive to the orientation of the two colliding nuclei as the number of collisions suffered by each participating quark inside the other colliding nucleus will be affected due to the available travel path according to their orientation. In a heavy-ion collision experiment, It is quite difficult to control the orientation of the two colliding nuclei. In our model, We have calculated the pseudorapidity density for central $U-U$ collisions by two ways : first we take average over all type of configuration for central collision and in second case we only take the tip-tip configuration by fixing both the angles $\theta_{1}$ and $\theta_{2}$ to zero and then calculate the $dn_{ch}/d\eta$ for central collisions. We tabulated our model results in Table. III along with IP-Glasma model result. Further we compare both model results with the experimental data. We find that IP-Glasma model as well as WQM provide a reasonable agreement with the data.  

\begin{table}
\begin{center}
\caption{The pseudorapidity distribution of $U-U$ in minimum bias and tip-tip at $\sqrt{s_{NN}}=193$ GeV by our model compared with IP-Glasma. Experimental data is taken from reference~\cite{12}.} 
\begin{tabular}{cccc} \hline \hline 
                             
                   & Colliding Nuclei     & $(dn_{ch}/d\eta)_{\eta=0}^{central}$        &   $(dn_{ch}/d\eta)_{\eta=0}^{tip-tip}$               \\ 
  & &(0-5$\%$)&(0-5$\%$)\\
\hline 
Our Model            & $U-U$               & $797$             & $739$      \\
IP-Glasma Model~\cite{26}         & $U-U$               & $824$                 & $815$    \\ 
Experimental data~\cite{12}       & $U-U$                & 830.4$\pm$67.8 (0-5$\%$)      &                       \\       \hline

\end{tabular}
\end{center}
\end{table} 
\subsection{Speed of Sound}
Transport properties are useful tool to quantify the behaviour of the matter created in the heavy-ion collisions~\cite{75}. Recently the data on collective velocity obtained from RHIC and LHC experiments indicate that a perfect fluid like system has been created in these collisions which is in contradiction to earlier prediction of the creation of an ideal QGP gas. Speed of sound is an important transport coefficient which can possibly hint the nature of matter created in these collisions since in ideal gas square of speed of sound can only go upto $0.33$ in magnitude. However if the system is in liquid form then speed of sound can cross this limit (since the speed of sound is large in liquid in comparison to gas). To calculate the square speed of sound $c_{s}^{2}$ from the pseudorapidity distribution, we have used the prescription as given in Ref.~\cite{23}:
\begin{equation}
c_{s}^{2} = \frac{1}{3\sigma^{2}}\left [\sqrt{16\ln ^{2}\left (\frac{\sqrt{s_{NN}}}{2m_{N}}  \right )+9\sigma^{^{2}} }-4\ln \left (\frac{\sqrt{s_{NN}}}{2m_{N}}  \right )\right].
\end{equation}\\
where $\sigma$ is rapidity distribution width and $m_{N}$ denotes the mass of a proton. The value of $\sigma$ is taken from Refs.~\cite{73,74} for different colliding nuclei $Cu-Cu$, $Au-Au$ and $Pb-Pb$ at different RHIC and LHC energies respectively. Our fitting function for pseudorapidity distribution with respect to pseudorapidity is little bit different from the function used in Ref.~\cite{23}. However, the role of $\sigma$ is similar to the Ref.~\cite{23} which is to provide a width to the pseudorapidity distribution and thus one can use the prescription to calculate $c_{s}^{2}$. 
\begin{figure}
\includegraphics[scale=0.55]{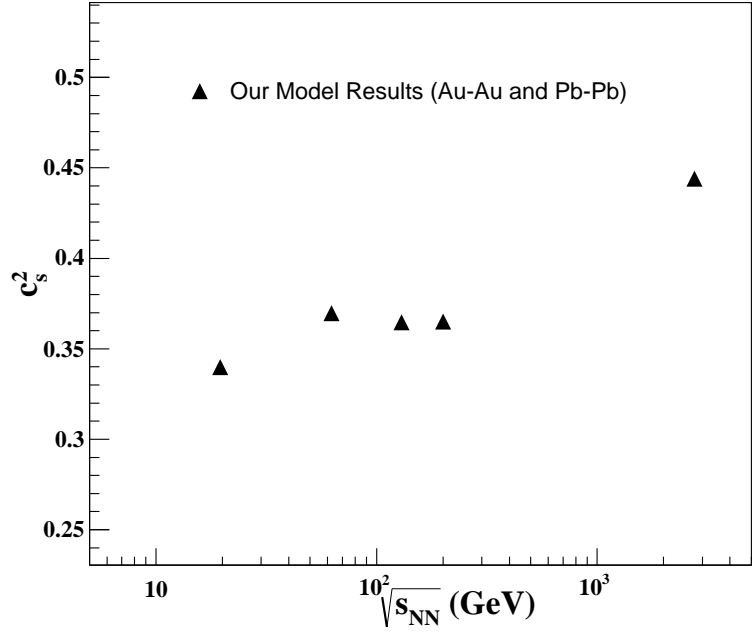}
\caption{Variation of speed of sound ($c_{s}^{2}$) with $\sqrt{s_{NN}}$ for $Au-Au$ and $Pb-Pb$ at RHIC and LHC energy respectively.}
\end{figure}

\begin{figure}
\includegraphics[scale=0.55]{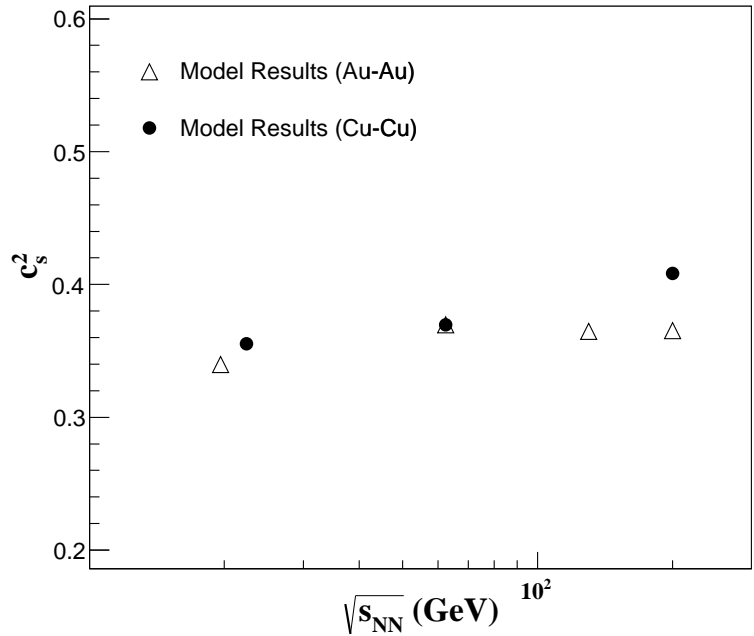}
\caption{Variation of $c_{s}^{2}$ in $Cu-Cu$ and $Au-Au$ with $\sqrt{s_{NN}}$ at RHIC energy.}
\end{figure}

In Fig. 14, we show the variation of square of speed of sound ($c_{s}^{2}$) in $Au-Au$ and in $Pb-Pb$ systems at different RHIC and LHC energy. Here, we find that the value of $c_{s}^{2}$ goes from $0.33$ to $0.37$ in $Au-Au$ collisions and its value in $Pb-Pb$ collisions at LHC energy is found to be $0.435$. From this we infer that the systems created in $Au-Au$ and $Pb-Pb$ collisions is not behaving like a ideal hadronic gas for which maximum value of $c_{s}^{2}$ can go upto $0.33$. In these collisions the higher values of $c_{s}^{2}$ indicate the formation of a medium behaving as a viscous fluid in contradiction to ideal gas. Further the larger value at LHC energy shows that the medium formed at this energy is more viscous in comparison to the medium formed at RHIC energy. In Fig. 15, the variation of square of speed of sound in $Au-Au$ and $Cu-Cu$ systems is shown. We see that in $Cu-Cu$ system $c_{s}^{2}$ is more than in $Au-Au$ systems at some energies and at some energies $c_{s}^{2}$ have almost same value as in the case of $Au-Au$ collisions. The medium created in smaller colliding system like $Cu-Cu$, the finite size effects have important role which causes fluctuation in the mean value of $c_{s}^{2}$ ~\cite{76} which makes speed of sound a bit random. 

In summary, we have analyzed the pseudorapidity density and multiplicity data of charged hadrons with respect to different control parameters in the view of a new wounded quark model. We proposed a new parametrization to calculate particle production in $p$-$p$ collisions and then extend them for $A$-$A$ collisions using wounded quark picture in which a quark suffers multiple collisions before hadronization. Different type of nuclear density function has been implemented in this model to calculate the particle production in symmetric, asymmetric and deformed nuclei collisions. We first demonstrate the variation of total mean multiplicity and pseudorapidity density of charged hadrons in symmetric $Pb$-$Pb$ collisions with respect to centrality and compare them with various theoretical models. Further particle mean multiplicities in $d$-$Au$ and $p$-$Pb$ collisions has been calculated within WQM and their comparison with wounded nucleon as well as colour glass condensate model is shown. Later we have calculated $dn_{ch}/d\eta$ of charged hadrons at midrapidity in asymmetric collisions of $Cu$-$Au$ as well as deformed $U$-$U$ nuclei collisions and compared them with the corresponding results of the other models. The agreement between the data and WQM results suggest that the quark picture more suits as a particle production mechanism in ultrarelativistic heavy ion collisions. We have also use WQM to calculate the speed of sound in heavy ion collisions.

\section{Acknowledgments}
\noindent 
O S K Chaturvedi and PKS is grateful to Council of Scientific and Industrial Research (CSIR), New Delhi for providing a research grant. One of the authors AK would like to acknowledge the financial support from DST-SERB project of Prof. B. Mohanty.

%\newpage


\begin{thebibliography}{50}

\bibitem{1} C. P. Singh, Phys. Rep. {\bf 236}, 147 (1993).

\bibitem{2} P. B.-Munzinger, K. Redlich, J. Stachel, arXiv:nucl-th/030401.

\bibitem{3} I. M. Dremin, J. W. Gary, Phys. Rep. {\bf 349}, 301 (2001).

\bibitem{4} ALICE Collaboration,  J. Adam et al., arXiv:nucl-ex/1509.07541.

\bibitem{5} ALICE Collaboration, E. Abbas et. al., Phys. Lett. B {\bf 726}, 610-622 (2013).

\bibitem{6} ALICE Collaboration, B. Abelev et al., Phys. Rev. C {\bf 88}, 044910 (2013). 

\bibitem{7} ALICE Collaboration, B. Abelev et al., Phys. Lett. B {\bf 754}, 373-385 (2016).

\bibitem{8} ALICE Collaboration, K. Aamodt et al., Phys. Rev. Lett. {\bf 106}, 032301 (2011).

\bibitem{9} PHOBOS Collaboration, B. B. Back et al., Phys. Rev. C {\bf 72}, 031901(R) (2005).

\bibitem{10} ALICE Collaboration, B. Abelev et al.,  Phys. Lett. B {\bf 719}, 29-41 (2013).

\bibitem{11} PHOBOS Collaboration, B. B. Back et al., Phys. Rev. Lett. {\bf 93}, 082301 (2004).

\bibitem{12} PHENIX Collaboration, A. Adare et al., Phys. Rev. C {\bf 93}, 024901 (2016).

\bibitem{13} ALICE Collaboration, J. Adam et al., Phys. Rev. C {\bf 91}, 064905 (2015).

\bibitem{14} ALICE Collaboration, B. Abelev et al., arXiv:nucl-ex/1512.06104.

\bibitem{15} ALICE Collaboration, B. Abelev et al., Phys. Lett. B {\bf 728}, 25-38 (2014).

\bibitem{16} A. Toia ( for ALICE Collaboration), J. Phys. G: Nucl. Part. Phys. {\bf 38}, 124007 (2011).

\bibitem{17} A. Iordanova, (for PHENIX Collaboration), J. Phys.: Conf. Series {\bf 458}, 012004 (2013).

\bibitem{18} C. Oppedisano (for ALICE Collaboration), J. Phys.: Conf. Series {\bf 455}, 012008 (2013).

\bibitem{19} R. S. Hollis (for PHENIX Collaboration), Nucl. Phys. A {\bf 904-905}, 507c-510c (2013).

\bibitem{20} PHOBOS Collaboration, B. B. Back et al., nucl-ex/0301017.
\bibitem{21} S. Jeon, V. T. Pop, M. Bleicher, Phys. Rev. C {\bf 69}, 044904 (2004).
\bibitem{22} PHOBOS Collaboration, B. B. Back et al., Phys. Rev. Lett. {\bf 88}, 022302 (2002).

\bibitem{23} L.-N. Gao and F.-H. Liu, Advances in High Energy Physics, {\bf 2015}, 184713, 2015.

\bibitem{24} A. N. Mishra, P. Sahoo, P. Pareek, N. K. Behera, R. Sahoo, B. K. Nandi, arXiv:hep-ph/1505.00700.

\bibitem{25} Y. H. Masaru, Hongo and T. Hirano, Phys. Rev. C {\bf 72}, 021903(R) (2014).

\bibitem{26} B. Schenke, P. Tribedy, and R. Venugopalan, Phys. Rev. C {\bf 89}, 064908 (2014).
 
\bibitem{27} H. Masui, B. Mohanty, N. Xu,  Phys. Lett. B {\bf 679}, 440-444 (2009).

\bibitem{28} H. Wang and P. Sorensen (STAR Collaboration), arXiv:nucl-ex/1406.7522.

\bibitem{29} U. Heinz and A. Kuhlman, arXiv:nucl-th/0411054.

\bibitem{30} A. Kuhlman and U. Heinz, arXiv:nucl-th/0506088.

\bibitem{31} A. Bzdak and V. Skokov, Phys. Rev. Lett. {\bf 111}, 182301 (2013).

\bibitem{32} J. L. Albacete, A. Dumitru, H. Fujii, Y. Nara, Nucl. Phys. A {\bf 897}, 1-27 (2013).

\bibitem{33} J. L. Albacete, A. Dumitru, arXiv:hep-ph/1011.5161.

\bibitem{34} A. Bialas, W. Czyz, and W. Furmanski, Acta Phys. Polon. B {\bf 8}, 585 (1977).
\bibitem{35} A. Bialas, K. Fialkowski, W. Slominski, and M. Zielinski, Acta Phys. Polon. B {\bf 8}, 855 (1977).
\bibitem{36} A. Bialas and W. Czyz, Acta Phys. Polon. B {\bf 10}, 831 (1979).
\bibitem{37} V. V. Anisovich, Yu. M. Shabelski, and V. M. Shekhter, Nucl. Phys. B {\bf 133}, 477 (1978).

\bibitem{38} A. Bialas and A. Bzdak, Phys. Rev. C {\bf 77}, 034908 (2008).

\bibitem{39} R. Engel, J. Ranft, S. Roesler, Phys. Rev. D {\bf 52}, 3 (1995).

\bibitem{40} M. Mitrovski, T. Schuster, G. Graf, H. Petersen and M. Bleicher, Phys. Rev. C {\bf 79}, 044901 (2009).

\bibitem{41} P. Bozek, W. Broniowski, and M. Rybczynski, Phys. Rev. C {\bf 94}, 014902 (2016).

\bibitem{42} H. Song and U. W. Heinz, Phys. Rev. C {\bf 78}, 024902 (2016).


\bibitem{43} F. W. Bopp, A. Capella, J. Ranft, J. Tran Thanh Van, Z. Phys. C {\bf 51}, 99-105 (1990).

\bibitem{44} M. A. Braun, F. del Moral and C. Pajares,  Phys. Rev. C {\bf 65}, 024907 (2002).

\bibitem{45} S. Eremin and S. Voloshin, Phys. Rev. C {\bf 67}, 064905 (2003).

\bibitem{46} P. K. Netrakanti and B. Mohanty, Phys. Rev. C {\bf 70}, 027901 (2004).

\bibitem{47} A. Dumitru, D. E. Kharzeev, E. M. Levin and Y. Narag,  arXiv:hep-ph/1111.3031.

\bibitem{48} G. G. Barnafoldi, J. Barrette, M. Gyulassy, P. Levai, and V. Topor Pop  arXiv:hep-ph/1111.3646.

\bibitem{49} R. A. Lacey, P. Liu, N. Magdy, M. Csand, B. Schweid, N. N. Ajitanand, J. Alexander, and R. Pak, arXiv:nucl-ex/1601.06001.
\bibitem{50} L. Zheng and Z. Yin, Eur. Phys. J. A {\bf 52}, 45 (2016).
%\bibitem{51} P. Bozek, W. Broniowski and M. Rybczynski, Phys. Rev. C {\bf 94}, 014902 (2016).
\bibitem{52} J. T. Mitchell, D. V. Perepelitsa, M. J. Tannenbaum, and P. W. Stankus, Phys. Rev. C {\bf 93}, 054910 (2016).
\bibitem{53} C. Loizides, Phys. Rev. C {\bf 94}, 024914 (2016).


\bibitem{54} G. Arnison et al., Phys. Lett. B {\bf 123}, 108 (1983).

\bibitem{55} C. Albajar et al., Nucl. Phys. B {\bf 335}, 261 (1990).

\bibitem{56} R. E. Ansorge et al., Z. Phys. C {\bf 43}, 357 (1989).

\bibitem{57} G. J. Alner et al., Phys. Lett. B {\bf 160}, 193 (1985).

\bibitem{58} G. J. Alner et al., Phys. Rep. {\bf 154}, 247 (1987).

\bibitem{59} F. Abe et al., Phys. Rev. D {\bf 41}, 2330 (1990).

\bibitem{60} W. Thome et al., Nucl. Phys. B {\bf 129}, 365 (1977).

\bibitem{61} J.-X. Sun, F.-H. Liu, E.-Q. Wang, Y. Sun, Z. Sun, Phys. Rev. C {\bf 83}, 014001 (2011).

\bibitem{62} R. Nouicer et al., J. Phys. G {\bf 30}, S1133 (2004).

\bibitem{63} B. I. Abelev et al., Phys. Rev. C {\bf 79}, 034909 (2009).

\bibitem{64} K. Aamodt et al., Eur. Phys. J. C {\bf 68}, 89 (2010).

\bibitem{65} K. Aamodt et al., Eur. Phys. J. C {\bf 65}, 111 (2010).

\bibitem{66} V. Khachatryan et al., J. High Energy Phys. {\bf 2010}, 02041 (2010).
\bibitem{67} V. Khachatryan et al., Phys. Rev. Lett. {\bf 105}, 022002 (2010).


\bibitem{68} C. Loizides, J. Nagle, P. Steinberg, arXiv:nucl-ex/1408.2549.

\bibitem{69} Q. Y. Shou et al.,  Phys. Lett. B {\bf 749}, 215-220 (2015).

\bibitem{70} C. P. Singh, M. Shyam, and S. K. Tuli,  Phys. Rev. C {\bf 40}, 1716 (1989).

\bibitem{71} C. P. Singh and M. Shyam, Phys. Lett. B {\bf 171}, 125 (1986). 

\bibitem{72} M. Shyam, C. P. Singh and S. K. Tuli, Phys. Lett. B {\bf 164} 189 (1985).

\bibitem{73} A. Kumar, B. K. Singh, P. K. Srivastava, and C. P. Singh, Eur. Phys. J. Plus {\bf 2013}, 128, (2013).

\bibitem{74} A. Kumar, P. K. Srivastava, B. K. Singh and C. P. Singh, Advances in High Energy Physics {\bf 2013}, 352180 (2013).


\bibitem{75} P. K. Srivastava and C. P. Singh, Phys. Rev. D {\bf 85}, 114016, (2012).

\bibitem{76} P. Ghosh, S. Ghosh, J. Mitra, N. Bera, Eur. J. Phys. {\bf 36}, 055046 (2015).







\end{thebibliography}
\end{document}